\documentclass[aps,prd,groupedaddress,preprintnumbers,longbibliography,twocolumn,showkeys,showpacs]{revtex4-1}
\usepackage{commath}
\usepackage{mathtools}
\usepackage{amssymb,amsfonts}
\usepackage{amsmath}
\usepackage[english]{babel}
\usepackage{mathrsfs}
\usepackage{hyperref}
\usepackage{graphicx}
\usepackage{natbib}
\usepackage[usenames,dvipsnames]{xcolor}
\graphicspath{{Figures/}}
\newtheorem{Theorem}{Theorem}
\newtheorem{Definition}{Definition}
\newtheorem{Lemma}{Lemma}

\begin{document}
\title{A Quasigroup Approach for Conservation Laws in Asymptotically Flat Spacetimes}

\author{Alfonso Zack Robles \ }
   \email[]{Alfonso.Robles@alumno.udg.mx} 
\author{Alexander I Nesterov}
  \email{nesterov@cencar.udg.mx}
\author{Claudia Moreno}
  \email{claudia.moreno@cucei.udg.mx}
\affiliation{Departamento de F{\'\i}sica, CUCEI, Universidad de Guadalajara,
	Av. Revoluci\'on 1500, Guadalajara, CP 44420, Jalisco, M\'exico}


\date{\today}

\begin{abstract}
In the framework of the quasigroup approach to conservation laws in general relativity, we show how the infinite-parametric Newman--Unti group of asymptotic symmetries can be reduced to the Poincar\'e quasigroup. We compute Noether's charges associated with any element of the Poincar\'e quasialgebra. The integral conserved quantities of energy momentum and angular momentum, being linear on generators of the Poincar\'e quasigroup,  are identically equal to zero in Minkowski spacetime. We present a definition of the angular momentum free of the supertranslation ambiguity. We provide an appropriate notion of intrinsic angular momentum and a description of the mass reference frame's center at future null infinity. Finally, in the center of mass reference frame, the momentum and angular momentum are defined by the Komar expression.
\end{abstract}



\maketitle

\section{INTRODUCTION} 

The general covariance of the Einstein equations results in differential conservation laws related to the field equations. From~Noether's theorem, it follows that for an arbitrary diffeomorphism generated by the vector field $\xi$, the~invariance of the Lagrangian for the Einstein equations leads to the conservation laws of the form, ${\frak J}^{\mu}(\xi)_{,\mu}=0$, where the vector density ${\frak J}^{\mu}$ is defined as ${\frak J}^{\mu}(\xi)={\frak h}^{\mu\nu}_{,\nu}$, with~${\frak h}^{\mu\nu}=-{\frak h}^{\nu\mu}$ being the  superpotential
, which is constructed from the densities of spin, bispin, vector field $\xi$,
and its derivatives~\cite{1,2,3}.

For the Einstein--Hilbert action, we have a simple expression for the Noether current,
\begin{equation}
{\frak J}^\mu= -\frac{1}{4\pi}\left(\sqrt{-g}\xi^{[\mu;\nu]} \right)_{,\nu},
\label{K} 
\end{equation} 
which was first obtained by Komar~\cite{4}. Komar's expression provides a fully satisfactory notion of the total mass in stationary, asymptotically flat spacetimes. It is worth noting that the normalization for energy momentum and angular momentum differs by a factor of 2, and~it is impossible simply to ``renormalize'' ${\frak J}^\mu$. The~resolution of this problem was pointed out in~\cite{5}. One needs to add to the Einstein--Hilbert action a surface term and apply Noether's theorem to the total~action.

In Minkowski space, the~existence of the Poincar\'e group and corresponding Killing vectors leads to the definitions of total momentum and total angular momentum. However, the~situation is more complicated in curved spacetime, even for an isolated system with a vanishing curvature tensor at infinity. While we have a well-defined energy momentum, there is no accordance for the notion of the angular momentum or center of mass~\cite{52}. A~major difficulty in defining the angular momentum is that the group of asymptotic symmetries is infinite-parametric. Although~the asymptotic symmetry group has a unique translation subgroup, there is no canonical Lorentz subgroup~\cite{8,13,35,40}. The~last one emerges as a factor group of the asymptotic symmetries group by the infinite-dimensional subgroup of supertranslations. Therefore, there does not exist a canonical way of choosing the Poincar\'e group as a subgroup of the group of asymptotic symmetries. There are too many Poincar\'e subgroups, one for each supertranslation, which is not translation \cite{10}.

These circumstances generate the main difficulties in the numerous attempts to find the correct definition of angular momentum in curved spacetime~\cite{11,12,39,43}. Most definitions suffer from the supertranslation ambiguities~\cite{6,7,8,9,10,11,12,13,14} (see~\cite{15} for review). The~origin-dependence of the angular momentum causes additional difficulties. Hitherto, no satisfactory way of resolving these problems has been~found.

In flat spacetime, a~total angular momentum can be written as $J^{\mu \nu} =M^{\mu \nu} + S^{\mu \nu}$, where the first term, $M^{\mu \nu} = X^{\mu }P^{\nu} - P^{\mu }X^{\nu} $, is the angular momentum,  with~$X^\mu$ and $P^\nu$ being the position vector and momentum of the particle, respectively, and~$ S^{\mu \nu}$ is the intrinsic angular momentum~\cite{9}.

In Minkowski spacetime, one can find a particular trajectory, describing the center of mass motion and having the property that the four-momentum is aligned with the observer's four-velocity~\cite{1,2}. Thus, if~one wants to generalize this concept to general relativity, then the task would be to find a worldline with similar properties in curved spacetime~\cite{16,17,18,19}. An~alternative approach is to define a preferred section at $\mathscr{I}^+$, which can be associated with the rest reference frame. This approach was developed in~\cite{20,21}, where the so-called "nice sections" were introduced to study the asymptotic fields of an isolated~system.

In our paper, the~quasigroup approach to the conservation laws developed in~\cite{2,22,23,24} is applied to asymptotically flat spacetime. The~Poincar\'e quasigroup at future null infinity ($\mathscr{I}^{+}$)
is introduced and compared with other definitions of asymptotic symmetries that have appeared in the literature.  We define the complex Noether charge
associated with any element of the Poincar\'e quasialgebra. It may be regarded as a form of linkages by Tamburino and Winicour~\cite{6,7} but with the new gauge conditions for asymptotic~symmetries. 

We present new definitions of the center of mass and intrinsic angular momentum using the available tools on asymptotically flat spacetimes. Our approach is based on the fact that the intrinsic angular momentum is invariant under the Lorentz boosts~\cite{26}. It is shown that the momentum and angular momentum are defined by the Komar expression in the center of the mass reference~frame.

The metric's signature is $(+, -,-,-)$ throughout this paper, with lower case Greek letters ranging and summing from zero to~three. 

This paper is organized as follows. In~Section~\ref{sec2}, we review the key ideas and tools that are indispensable for further discussions. In~particular, we discuss the non-associative generalization of the transformation groups--quasigroups of transformations and the structure of the group of asymptotic symmetries at future null infinity. In~Section~\ref{sec3}, we explore the reduction of the Newman--Unti (NU) to the Poincar\'e quasigroup. In~Section~\ref{sec4}, we present conserved quantities at future null infinity based on linkages introduced by Tamburino and Winicour. In~Section~\ref{COM}, we define a cut of $\mathscr{I}^{+}$ associated with the notion center-of-mass reference frame of isolated systems and show that the Komar expression gives the intrinsic angular momentum. Finally, we give our conclusion Section~\ref{sec6}, 
summarizing the obtained results and discuss possible generalizations of our approach. In~the appendices, the~details of our calculations are~presented.

\section{Mathematical~Background}\label{sec2}

\subsection{Quasigroups of Transformations}

The definition of the quasigroup of transformations was first given by
Batalin~\cite{28}. Below,~we outline the main facts from the theory of the
smooth quasigroups of~transformations.

Let $\frak M$ be an n-dimensional manifold and the continuous law
of transformation is given by $x'=T_{a}x$ , $x\in {\frak M}$, where
$\{a^i\}$ is the set of real parameters, $i=1,2,\dots,r$. The~set of
transformations $\{T_{a}\}$ forms an $r$-parametric quasigroup of
transformations (with right action on ${\frak M}$): 
\begin{enumerate}
	\item If there exists a
unit element that is common for all $x^\alpha$ and corresponds to
$a^i=0:$, then $T_a x|_{a=0}=x$;
	\item If the modified composition law holds,
\begin{equation}
			T_aT_b x=T_{\varphi(b,a;x)} x;
		\end{equation}
	\item If the left and right units coincide,
		\begin{align*}
			\varphi(a,0;x)=a, \quad \varphi(0,b;x)=b;
		\end{align*}
	\item If the modified law of associativity is satisfied,
		\begin{align*}
		\varphi(\varphi(a,b;x),c;x)=\varphi(a,\varphi(b,c;T_a x);x);
		\end{align*}
	\item If the transformation inverse to $T_a$ exists, $x=T^{-1}_a x'$.
\end{enumerate} 

A group is a set with a binary operation that is closed, associative, has an identity element, and~every element has an inverse. In~contrast, a~quasigroup is a more generalized algebraic structure that only requires the operation to be~closed.
        
The quasigroup structure ``falls short'' of being a group in one specific way: associativity is modified. In~a group the definition is,
        \begin{equation}
            (T_{a} \cdot T_{b}) \cdot T_{c} = T_{a} \cdot (T_{b} \cdot T_{c}),    
        \end{equation}
 and a quasigroup,
\begin{align}
    (T_{a} \cdot T_{b}) \cdot T_{c} = T_{a} \cdot (T_{b} \cdot T_{c'}),
\end{align}
where $c'$ depends on $T_{a}$.
        
A more intuitive understanding is that the sequence in which operations are grouped is significant, as~implementing the initial transformation alters the foundational context for subsequent transformations~\cite{28,31}.
        
By the other side, the~generators of infinitesimal transformations,
\begin{equation}
\Gamma_i=(\partial(T_ax)^\alpha/\partial a^i)|_{a=0}
\partial/\partial x^\alpha \equiv R^\alpha_i\partial/\partial x^\alpha,
\end{equation}
form quasialgebra and obey commutation relations:
\begin{equation}
[\Gamma_i, \Gamma_j] =C^p_{ij}(x)\Gamma_p.
\label{eq1}
\end{equation}

Here, $C^p_{ij}(x)$ denotes the structure functions satisfying the modified
Jacobi identity:
\begin{align}
C^p_{ij,\alpha}R^\alpha_k + C^p_{jk,\alpha}R^\alpha_i
+C^p_{ki,\alpha}R^\alpha_j&\nonumber\\
+ C^l_{ij}C^p_{kl} + C^l_{jk}C^p_{il} + C^l_{ki}C^p_{jl}&=0 .
\label{eq2}
\end{align}

\begin{Theorem}
Let 
 the given functions $R^\alpha_i, \;C^p_{kj}$
obey Equations~(\ref{eq1}) and (\ref{eq2}); then, locally, the quasigroup
of transformations is reconstructed as the solution of a set of differential
equations:
\begin{align}
& \frac{\partial\tilde x^\alpha}{\partial a^i}
= R^\alpha_j(\tilde x)\lambda^j_i(a;x),\,
\tilde x^\alpha(0)=x^\alpha, \label{eq3} \\
& \frac{\partial\lambda^i_j}{\partial a^p}
-\frac{\partial\lambda^i_p}{\partial a^j}
+ C^i_{mn}(\tilde x)\lambda^m_p\lambda^n_j=0, \label{eq4}\\
&\lambda^i_j(0;x)=\delta^i_j . \nonumber
\end{align}
\end{Theorem}

Equation~(\ref{eq3}) is an analog of the Lie equation, and~Equation~(\ref{eq4}) is
the generalized Maurer--Cartan equation~\cite{28}.

\begin{Definition}
The  Poincar\'e quasigroup is defined as a semi-direct product of the Lorentz quasigroup and the group of~translations.

Poincar\'e quasialgebra is described by the following set of commutation relations:
\begin{align}
[ T_a ,T_b ]&=0,\nonumber\\ 
[T_a,L_B]&=C^b_{aB} (x)T_b,\nonumber\\ \quad 
\left [ L_A ,L_B \right]&= C^D_{AB}(x)L_D,
\end{align}
with $C^D_{AB},\;C^b_{aB}$  being the {\it structure functions}. The~last
commutation relations mean that the generators of the Lorentz quasigroup form a closed quasialgebra~\cite{24}. 
\end{Definition}

\subsection{NP Formalism }

This section outlines the main facts from the Newman--Penrose (NP) formalism indispensable for future discussions. We follow notations in~\cite{29,30,31}.

 At each point of spacetime, we introduce a null NP basis
$\{l,n,m,\bar m\}$ with the following non-vanishing scalar products:
$l\cdot n=1,~m\cdot \bar m =-1$. The~metric components are given by
\begin{align}
g^{\alpha\beta}=l^{\alpha}n^{\beta} + l^{\beta} n^{\alpha}
-m^{\alpha}\bar m^{\beta} - \bar{m}^{\beta} m^{\alpha},
\end{align}
where {\it bar} means complex~conjugate.

To introduce the coordinate system, we choose a one-parameter family of
two-dimensional space-like cross-sections of future null infinity, ${\mathscr{I}}^{+}$, which are labeled by a coordinate $u$. We assume that $l = \partial / \partial r$, being a future directed null vector, is tangent to the surface $u$ = const and a null vector $n$ is parallelly propagated along the $l$ congruence. An~affine parameter $r$ is normalized by the condition $l^\alpha \partial r/\partial x^\alpha =1$. Since the topology of ${\mathscr{I}}^{+}$ is $S^2\times R$, ``cuts'' of ${\mathscr{I}}^{+}$ can be labeled by a complex stereographic coordinate $\zeta = e^{i\varphi}\cot \theta/2$ \cite{32}.

From the null tetrad basis, complex Ricci rotation coefficients are defined
as follows:
\begin{flalign}
\rho & = l_{\alpha;\beta}m^\alpha \bar m^\beta,\qquad
\mu = -n_{\alpha;\beta}m^\alpha m^\beta,\\
\sigma & = l_{\alpha;\beta}m^\alpha m^\beta,\qquad
\lambda = -n_{\alpha;\beta}\bar m^\alpha \bar m^\beta,\\
\kappa & = l_{\alpha;\beta}m^\alpha l^\beta,\qquad
\nu = -n_{\alpha;\beta}\bar m^\alpha n^\beta,\\
\tau & = l_{\alpha;\beta}m^\alpha n^\beta,\qquad
\pi = -n_{\alpha;\beta}\bar m^\alpha l^\beta,\\
\epsilon & =\frac{1}{2}(l_{\alpha;\beta}n^\alpha l^\beta
 -m_{\alpha;\beta}\bar m^\alpha l^\beta),\\
\gamma & =-\frac{1}{2}(n_{\alpha;\beta}l^\alpha n^\beta
 - \bar m_{\alpha;\beta} m^\alpha n^\beta),\\
\alpha & =\frac{1}{2}(l_{\alpha;\beta}n^\alpha \bar m^\beta
 - m_{\alpha;\beta}\bar m^\alpha \bar m^\beta),\\
\beta & =-\frac{1}{2}(n_{\alpha;\beta}l^\alpha m^\beta
 - \bar m_{\alpha;\beta} m^\alpha m^\beta).
\end{flalign}

The Riemann tensor is decomposed into its irreducible parts in such a way
that the corresponding tetrad components are labeled:
\begin{align}
\Psi_0 =& - C_{\alpha\beta\gamma\delta}l^\alpha m^\beta l^\gamma m^\delta
,\\
\Psi_1 =& - C_{\alpha\beta\gamma\delta}l^\alpha n^\beta l^\gamma m^\delta
,\\
\Psi_2 =& -\frac{1}{2} C_{\alpha\beta\gamma\delta}(l^\alpha n^\beta l^\gamma
n^\delta - l^\alpha n^\beta m^\gamma \bar m^\delta),\\
\Psi_3 =& - C_{\alpha\beta\gamma\delta}l^\alpha n^\beta \bar m^\gamma
n^\delta ,\\
\Psi_4 =& - C_{\alpha\beta\gamma\delta}n^\alpha \bar m^\beta n^\gamma
\bar m^\delta \label{Psi4} ,
\end{align}
where $C_{\alpha\beta\gamma\delta}$ is the Weyl~tensor.

Further, it is convenient to introduce the notation
\begin{align}
\begin{split}
D =l^{\alpha}\nabla_{\alpha}, \quad \Delta = n^{\alpha}\nabla_{\alpha}, \\
\delta = m^{\alpha}\nabla_{\alpha}, \quad \bar\delta =\bar m^{\alpha}\nabla_{\alpha}	,
\end{split}
\end{align}
for the projections of the covariant derivatives onto the null tetrad. Then, one can write
\begin{align}
D =~&\frac{\partial}{\partial r},\qquad
\Delta=\frac{\partial}{\partial u} +U\frac{\partial}{\partial r}
+ X^A\frac{\partial}{\partial x^A}, \nonumber \\
\delta=~&\omega\frac{\partial}{\partial r}
+ \xi^A\frac{\partial}{\partial x^A}, \quad A=2,3, \nonumber
\end{align}
for some  $U, X^A, \omega$, and $\xi^A$, where $x^2=\zeta$,  $x^3=\bar\zeta$ \cite{31,33}. 
 
In the following, employing the coordinate freedom, we will use the Bondi coordinates at $\mathscr{I}^+$. This choice of coordinates implies that a two-dimensional surface $S_\infty$, being obtained as a cut $u= \rm const$, is a two-sphere $S^2$ with the line element written as
\begin{align}
ds^2 = \frac{4 d\bar \zeta d \zeta}{1 + \bar \zeta  \zeta }.	
\end{align}

We choose the NP-basis at $\mathscr{I}^+$ as follows:
\begin{align}
\Delta^0~=~&\frac{\partial}{\partial u} , \\
\delta^0~=~& P\frac{\partial}{\partial \zeta} = -\frac{1}{\sqrt{2}}\left(\frac{\partial}{\partial \theta} +  \frac{i}{\sin \theta}\frac{\partial}{\partial \varphi}\right) ,
\end{align}
where $P= \zeta(1 + \zeta \bar \zeta)/(\sqrt{2} |\zeta|)$. Sign "$-$"
in the definition of $\delta^0$ is introduced to provide an agreement with the standard form of raising and lowering operators (see Appendix \ref{apendixA}).

The Weyl tensor components essential for future spin coefficients $\alpha$, $\beta$, $\lambda$, and~$\sigma$, asymptotically, are as follows~\cite{9,29,33}:
\begin{align}
	\alpha =~&\alpha^0 r^{-1}+\mathcal{O}\left(r^{-2}\right) ,\\
	\beta =~&\beta^0 r^{-1}+\mathcal{O}\left(r^{-2}\right), \\
	\lambda =~&\lambda^0 r^{-1}+\mathcal{O}\left(r^{-2}\right), \\
        \sigma=~&\sigma^{0} r^{-2}+\mathcal{O}\left(r^{-3}\right),\\
    \Psi_{0} =~&\Psi_{0}^{0} r^{-5}+\mathcal{O}\left(r^{-6}\right) ,\\
    \Psi_{1} =~&\Psi_{1}^{0} r^{-4}+\mathcal{O}\left(r^{-5}\right), \\
    \Psi_{2} =~&\Psi_{2}^{0} r^{-3}+\mathcal{O}\left(r^{-4}\right), \\
    \Psi_{3} =~&\Psi_{3}^{0} r^{-2}+\mathcal{O}\left(r^{-3}\right) ,\\
    \Psi_{4} =~&\Psi_{4}^{0} r^{-1}+\mathcal{O}\left(r^{-2}\right).
\end{align}

In Bondi coordinates, relationships between functions on $\mathscr{I}^+ $  are as follows~\cite{9,17,34}:
\begin{align}
    &\alpha^0 =\frac{\bar P}{2} \frac{\partial \ln P}{\partial \bar\zeta} , \\
    &\beta^0 = - \bar \alpha^0 , \quad  \lambda^0  =\dot{\bar \sigma}^0,
\end{align}
where $\lambda^{0}=\dot{\overline{\sigma}}^{0}$ is the news function; ``dot'' denotes the derivative $\partial/\partial u$; function $\lambda^{0}$ represents information emerging from the positive null infinity $\mathscr{I}^{+}$; and the shear $\sigma^{0}$ corresponds to the shear free parameter. With~the shear $h=h_{+}-ih_{\times}\propto\sigma^{0}$,~the news function is linked to its time-derivative $\dot{h}\propto\lambda^{0}$ gravitational wave polarizations $h$ from gauge-invariant perturbations derived by Thorne in~\cite{25,52}. 
 
 The raising and lowering operators $\eth$ and  $\bar\eth$, respectively, are defined by the following expressions:
\begin{align}
 	\eth \eta & = \delta \eta + s(\bar\alpha^0 - \beta)\eta , \\
 	\bar\eth  \eta & = \bar\delta \eta -  s(\bar\alpha^0 - \beta)\eta,
 	\label{eth}	
 \end{align}
 where $s$ is the spin weight of function $\eta$ {\cite{9,32}}.
 
 The Weyl tensor is as follows~\cite{17}: 
\begin{align} 
& \Psi^0_2 -\bar\Psi^0_2 = \bar\sigma^0\bar\lambda^0 - \sigma^0\lambda^0
+ \bar\eth^2\sigma^0 - \eth^2\bar\sigma^0, \label{Psi2} \\
& \Psi^0_3 =  - \eth\lambda^0 , \label{W1a}\\
& \Psi^0_4 = -\dot\lambda^0 .
 \label{W1b}
\end{align}

The mass aspect function~\cite{21,31,35},
\begin{align}
	\Psi = \Psi^0_2 + \sigma^0\lambda^0+  \eth^2\bar\sigma^0,
\end{align}
satisfies the reality condition $\Psi = \bar \Psi$.

The evolution equations (Bianci identities) have the following asymptotic form:
\begin{align}\label{W2}
\dot\Psi^0_0 =~ &\eth \Psi^0_1 + 3 \sigma^0 \Psi^0_2, \\
\dot\Psi^0_1 =~& \eth \Psi^0_2 + 2 \sigma^0 \Psi^0_3, \\
\dot\Psi^0_2 =~& \eth \Psi^0_3 +  \sigma^0 \Psi^0_4.
\label{W2b}
\end{align}

The relations in Equations \eqref{W1a}--\eqref{W2b} have the opposite sign in comparison with similar expressions in Ref.~\cite{29}. This is due to the difference in the sign of $\delta^0$. In~\cite{29}, it is defined as $\delta^0 = - P\partial_\zeta$.

\section{Asymptotic Symmetries and the Poincar\'e~Quasigroup}\label{sec3}

The asymptotic symmetries of the asymptotically flat spacetime at future null infinity are described by the infinite-parametric Newman--Unti (NU) group~\cite{29,33,36,37}. The~latter is defined by the transformation $\mathscr{I}^+ \rightarrow \mathscr{I}^+$, with the form
\begin{align}
u \rightarrow u' =~& f(u,\zeta,\bar \zeta) ,\quad
{\partial f}/{\partial u}> 0, \nonumber \\ 
\zeta \rightarrow \zeta' =~&
(\alpha\zeta+\beta)/(\gamma\zeta+\delta),
\quad \alpha\delta -\beta\gamma = 1, \nonumber
\end{align}
where $f(u,\zeta,\bar \zeta)$ is an arbitrary~function. 

The infinite-dimensional Bondi--Metzner--Sachs (BMS) group preserving strong conformal geometry is a subgroup of the NU group. The~BMS group is defined as follows~\cite{16,21,35,39,40}:
\begin{align}
&u \rightarrow u' = K(\zeta,\bar \zeta)(u+a(\zeta,\bar \zeta)), \\
&\zeta \rightarrow \zeta' = (\alpha\zeta+\beta)/(\gamma\zeta+\delta), \\
& \alpha\delta - \beta\gamma = 1, 
\end{align}
where $\alpha$,~$\beta$,~$\gamma$, and~$\delta$ are complex constants, and~$a(\zeta,\bar \zeta)$ is an arbitrary regular function on $S^2$; moreover,~\begin{align}
	 K(\zeta,\bar \zeta)=\frac{(1+\zeta\bar \zeta)}{(|\alpha\zeta+\beta|^2 +
|\gamma\zeta+\delta|^2)}.
\end{align}

The infinite-parameter normal subgroup of
the BMS group,
\begin{align}
	\zeta' = \zeta ,\quad u' = u+a(\zeta,\bar \zeta),
	\label{ST}
\end{align}
is called the subgroup of {supertranslations} and contains a
four-parameter normal translation subgroup:
\begin{align}
	a=\frac{p+q\zeta + \bar q \bar\zeta + g \zeta\bar\zeta}{1+\zeta\bar\zeta}, \quad \Im p = \Im g =0.
	\label{T1}
\end{align}

The quotient (factor) group of the BMS group by the supertranslations consists 
of the conformal transformations $S^2 \rightarrow S^2$, and it is isomorphic
to the proper orthochronous Lorentz group.

Since the BMS group is the semi-direct product of the Lorentz group and supertranslation's group, there is no canonical way to embed the Poincar\'e group in the BMS group. One has an infinite number of alternatives to extract the Poincar\'e group. However, at~least in Minkowski spacetime, one can elucidate which additional structure on $\mathscr{I}^+$ the Poincar\'e group preserves.  It turns out to be that the Poincar\'e group transforms the so-called good cuts~\cite{41}, cuts with vanishing shear $\sigma^0$.

This can be easily seen by considering the transformation of shear under supertranslation Equation \eqref{ST}. We obtain  $\sigma^0 \rightarrow \sigma'^0 = \sigma^0 - \eth^2 a$, where $\eth$ is  the ``eth'' operator on ${\mathscr{I}}^{+}$. Thus, a~supertranslation transforms a good cut to a bad cut---a cut with nonvanishing shear. If~we impose the condition $ \sigma'^0 =  \sigma^0 = 0$, we obtain $\eth^2 a =0$. The~solution of this equation with  a real $a$ yields a four-parametric subgroup of translations defined by Equation \eqref{T1}.

\subsection{Reduction of the NU Group to the Poincar\'e~Quasigroup}

The infinitesimal NU group is obtained from the asymptotic Killing
equations~\cite{36}:
\begin{align}
&\pounds_\xi g_{\mu\nu}=\xi_{\mu;\nu} +\xi_{\nu;\mu}=0(r^{-n}),\\
&l^{\nu}\pounds_\xi g_{\mu\nu}=(\xi_{\mu;\nu}
+\xi_{\nu;\mu})l^{\nu}=Q(u,\zeta,\bar\zeta)l_\mu,
\label{NU_alg}
\end{align}
where $Q(u,\zeta,\bar\zeta)$ is an arbitrary function on $\mathscr{I}^{+}$.

One can write a general element of NU-algebra as follows:
\begin{align}
&\xi=B(u,\zeta,\bar\zeta)\Delta^0 +
C(u,\zeta,\bar\zeta)\bar\delta^0 + \bar C(u,\zeta,\bar\zeta) \delta^0, \nonumber \\
&\eth C=0,
\label{eqG}
\end{align}
where $\eth$, $\Delta^0 $, and $\delta^0 $ are the standard NP
operators, and ``eth'', $\Delta$, and  $\delta$ are restricted on $\mathscr{I}^+$. 

The generators of the four-parameter translation subgroup are given by
\begin{align}
\xi_a=B_a(\zeta,\bar\zeta)\Delta^0, 
\end{align}
where the function $B_a$ is assumed to be a real function and is the solution of the following equation:
\begin{align}
	\eth^2 B_a=0, \quad \Im B_a =0 \quad (a = 0,1,2,3).
\end{align}

The generators of the ``Lorentz group'' are determined as follows:
\begin{align}\label{Q}
&\xi_A= B_A(u,\zeta,\bar\zeta)\Delta^0
+C_A\bar\delta^0+ \bar C_A \delta^0,  \\
&\eth C_A=0, \quad (A = 1,2 \dots 6),
\label{Q1}
\end{align}
where  $B_A(u,\zeta,\bar\zeta)$ is an arbitrary real~function.

The generators of the NU group obey the commutation relations:
\begin{align}\label{CR1}
& [ \xi_a ,\xi_b ]=0, \quad
[\xi_a,\xi_B]=C^b_{aB}(u,\zeta,\bar\zeta)\xi_b,  \\ 
&\left [ \xi_A
,\xi_B \right] = C^D_{AB}(u,\zeta,\bar\zeta)\xi_D, 
\label{CR}
\end{align}
where $C^b_{aB}$ and $C^D_{AB}$ are the structure functions. This points out that the NU group is a {\it quasigroup} with the closed Lorentz~quasialgebra.

To reduce the NU group to the particular Poincar\'e quasigroup, one needs to impose the 
constraints on a function $B_A(u,\zeta,\bar\zeta)$ and, thus, fix the supertranslational 
ambiguity in the definition of the Lorentz~quasigroup. 

In our approach, we use the fact that a group of isometries transforms an arbitrary geodesic to a geodesic one, and~the Killing vectors satisfy the geodesic deviation equation for any geodesic~\cite{23,24}. In~the construction below, only {\em null} geodesics passing inward are transformed to geodesics under the transformations of the Poincar\'e quasigroup. Instead of using the approximate Killing equations, we propagate the asymptotic generators $\xi$ defined on $\mathscr{I}^{+}$ inward along the null surface $\Gamma$ intersecting $\mathscr{I}^+$ in $\Sigma^{+}$ employing the geodesic deviation equation.
\begin{equation}
\nabla^2_l\xi + R(\xi,l)l=0.
\label{Jac}
\end{equation}

Since the geodesic deviation equation is the second-order ordinary differential equation for obtaining the unique solution, we need to impose the initial conditions on the vector $\xi$ and its first derivatives on $\mathscr{I}^+$. We use the asymptotic Killing equations for determining~them. 

The key idea behind our approach is to use the geodesic deviation equation only for a  null geodesic congruence passing inward $\mathscr{I}^+$ to define the generators of the Poincare quasigroup. This implies not only that the Poincar\'e quasigroup transforms an arbitrary geodesic to a geodesic but~also that the null geodesics belong to the null congruence defined~above. 

\subsubsection{Minkowski~Spacetime}

We demonstrate our approach in Minkowski spacetime.  The~key idea is to reduce the NU 
group to the ten-parametric Poincar\'e group, imposing the appropriate conditions on an 
arbitrary function $B_A$, and~thus fixing the supertranslational freedom (see Refs.~\cite{22,23,24} for details).

Let us write the Killing vector as
\begin{align}
\xi = A D + B \Delta + \bar C \delta +  	C \bar  \delta.
\end{align}

Using the asymptotic expansion
\begin{align}
    A =& A_1 r + A_0 + A_{-1}r^{-1} + 0(r^{-2}), \\
    B =& B_1r + B_0 + B_{-1}r^{-1} + 0(r^{-2}), \\
    C =& C_1 r + C_0 + C_{-1}r^{-1} + 0(r^{-2}), 
\end{align}
we obtain the solution of the geodesic deviation equation in the following
form:
\begin{align}
	A_{-n} &= A_{-n} (u,\zeta, \bar \zeta),\nonumber\\ 
    C_{-n} &=  C_{-n} (u,\zeta, \bar \zeta),\nonumber\\ 
    B_{-n} &= 0 \quad (n\geq 1).
\end{align}

The explicit dependence of $A_{-n} (u,\zeta, \bar \zeta)$, and~$	C_{-n} (u,\zeta, \bar \zeta)$ is not important to studying the structure of asymptotic~symmetries.

To obtain the unique solution of the geodesic deviation equation, we have
to impose the conditions on functions $A,B,$ and $C$ and their first
derivatives at $\mathscr{I}^{+}$. This implies that $A_0, A_1, B_0,B_1,C_0,$ and $C_1$ should be determined. We adapt the asymptotic Killing equations to determine these coefficients:
\begin{align}
&\lim_{r\rightarrow\infty}l^\mu
l^\nu\pounds_{\xi}g_{\mu\nu}= 0,\quad \lim_{r\rightarrow\infty}m^\mu
n^\nu\pounds_{\xi}g_{\mu\nu}= 0, \nonumber \\
&\lim_{r\rightarrow\infty}m^\mu \bar
m^\nu\pounds_{\xi}g_{\mu\nu}= 0, \,
\lim_{r\rightarrow\infty}r m^\mu \bar
m^\nu\pounds_{\xi}g_{\mu\nu}= 0,\nonumber \\
&\lim_{r\rightarrow\infty}rl^\mu m^\nu\pounds_{\xi}g_{\mu\nu}= 0.
\end{align}

After some algebra, we obtain
\begin{align}\label{EqK1}
& A_0=- \frac{1}{2} (\eth \bar C_0 +\bar\eth  C_0 ) - \frac{B_0}{2} (\mu^0 + \bar \mu^0 ) 
\nonumber\\
&+ \frac{1}{2} (\tau^0 \bar C_1+ \bar \tau^0 C_1 ), \\
& A_1=- \frac{1}{2} (\eth \bar C_1 +\bar\eth  C_1 ),  \\
& B_1 =0, \\
&C_0 = - \eth B_0 + \sigma^0 \bar C_1,
\label{EqK2a} \\
& \eth C_0 = \eth \sigma^0 C_1 + \frac{\sigma^0}{2} (\eth \bar C_1 - \bar\eth  C_1 ), 
\label{EqK2b}\\
&\dot C_1 =0, \quad \eth C_1 =0, 
\label{EqK2}
\end{align}
where ``dot''  denotes the derivative with respect to the  retarded time $u$. Substituting $C_0$ from Equation \eqref{EqK2a} in Equation \eqref{EqK2b}, we get
\begin{align}\label{EqK3}
&\dot C_1 =0, \quad \eth C_1 =0, \\
&\eth^2  B_0  -\frac{\sigma^0}{2}(3 \eth\bar
C_1 - \bar \eth C_1) - \bar C_1 \eth\sigma^0 - C_1 \bar\eth\sigma^0 =0.
\label{EqK4}
\end{align}

A general solution of this system can be written as
\begin{align}
&B_0	= B_t + \eth \eta \bar C _1+ \bar\eth \eta  C_1 + \frac{u - \eta}{2}( \eth \bar C_1  +\bar \eth  C_1   ),  \nonumber \\ 
&\eth^2 B_t=0, \quad \eth C _1=0,
\label{Zeta2}
\end{align}
where $\sigma^0= \eth^2 \eta$. 

The system of differential constraints, Equations \eqref{EqK3} and \eqref{EqK4}, is a unique one that determines functions $B_0$ and $C_1$ and restricts the NU group to a particular Poincar\'e group. Thus, in Minkowski spacetime, one can reduce the NU group to the Poincar\'e group even for ``bad'' cuts ($\sigma^0 \neq 0$).

Now, an arbitrary Killing vector at $\mathscr{I}^+$, $\xi^0 = \xi|_{\mathscr{I}^+}$, can be written as
\begin{equation}
\xi^0=B_0(u,\zeta,\bar \zeta)\Delta^0 + C_1(\zeta,\bar \zeta)\bar\delta^0+ \bar C_1(\zeta,\bar \zeta) \delta^0.
\label{EqKV}
\end{equation}

To specify the generators of the Poincar\'e group, one should impose the additional conditions on functions $B_0$ and $C_1$.

The generators of translations are,
\begin{align}
	l_a = B_a \Delta^0, \quad a=0,1,2,3,
\end{align}
with $\Im B_a =0$ and~$B_a$ denoting solutions of the differential equation $\eth^2 B_a =0$. With~real $B_a$, we obtain four independent solutions of this equation, yielding
\begin{align}
	l_a=(1, \sin\theta \cos \varphi, \sin\theta \sin\varphi, \cos \theta).
\end{align}
	
    The generators of boosts and rotations are given by
\begin{equation}
\xi_A =B_A\Delta^0 + C_A\bar\delta^0+ \bar C_A \delta^0,
\label{eqBR3}
\end{equation}
where $\eth C_A=0$ and
\begin{align}
B_A	=  \eth \eta \bar C_A + \bar\eth \eta  C_A + \frac{u - \eta}{2}( \eth \bar C_A  +\bar \eth  C_A   ),
\label{Seta3}
\end{align}
with $\Im B_A = \Im \eta =0$ and $\sigma^0 = \eth^2 \eta$.
Imposing additional conditions,
\begin{align}
	&\bar \eth C_A + \eth \bar C_A=0 \quad (\rm rotations), \\
	&\bar \eth C_A - \eth \bar C_A=0  \quad (\rm boosts),
\end{align}
we obtain six independent solutions of equation $\eth C_A=0$.  It is convenient to divide them into two groups, writing $C_A = \{ L_i,R_i \}$, where $L_i$ and $R_i$ describe the rotations and boosts, respectively.
We denote the generators of the complex Lorentz group as $\Gamma_A = \bar C_A \delta^0 = \xi_A{\partial_\zeta} $. Using the results of Ref.~\cite{37}, we obtain the generators of the boost and rotations. They are shown in Table~\ref{tab:placeholder}. 

\begin{table}[]
    \centering
    \begin{tabular}{|c|c|}
    \hline
    \textbf{Generator}  &  \boldmath{$ \Gamma_A=\xi_A\frac{\partial }{\partial \zeta}$}\\
    \hline \hline
 $\Gamma_1 =L_1$ &  $\frac{i}{2}(1- \zeta^2)\frac{\partial }{\partial \zeta} $\\
  $\Gamma_2 =L_2$ &  $-\frac{1}{2}(1+ \zeta^2)\frac{\partial }{\partial \zeta} $\\
    $\Gamma_3  =L_3$   &$i\zeta\frac{\partial }{\partial \zeta} $\\
 $\Gamma_4  =R_1$ &  $\frac{1}{2}(1- \zeta^2)\frac{\partial }{\partial \zeta} $\\
  $\Gamma_5  =R_2$ &  $\frac{i}{2}(1+ \zeta^2)\frac{\partial }{\partial \zeta} $\\
    $\Gamma_6  =R_3$ &  $\zeta\frac{\partial }{\partial \zeta} $\\
      \hline
    \end{tabular}
    \caption{Generators of the complex Lorentz group. (Note that $L_{1}$,~$L_{2}$, and $L_{3}$ are defined from $L_{x}$, $L_{y}$, and $L_{z}$ (see Appendix \ref{sec: Appendix B})).}
    \label{tab:placeholder}
\end{table}


Let us  introduce a complex vector $\xi_c$ at $\mathscr{I}^+$  such
that an arbitrary element of the infinitesimal NU group is written as $\xi= \xi_c +\bar \xi_c$, where
\begin{align}
	\xi_c =\xi_s\Delta^0 + \xi_1 \delta^0.
\end{align}

(Hereafter, we omit the index ``0'' in $\xi^0$.) We specify the vector $\xi_s$ as follows:
\begin{align}\label{K1}
&\xi_s=\xi_t + \eth\eta \xi_1 + \frac{u-\eta}{2}\eth\xi_1,\\
&\eth^2\xi_t^0 =0, \quad \bar\eth\xi_1=0,
\label{K2}
\end{align}
 where $\Im \xi^0_t = \Im \eta=0$ and $\eth^2 \eta = \sigma^0 $. Comparing this expression with Equation \eqref{EqKV}, we find $B_0 = \xi_s + \bar \xi_s $ and $C_1 = \bar \xi_1$.  

A straightforward computation shows that $\xi_s$ obeys the following differential equation:
\begin{align}
\begin{split}
		\eth^2 \xi_s =~& \frac{3}{2}\sigma^0\eth\xi_1 + \eth \sigma^0 \xi_1,  \\
	\bar \eth \xi_1 =~&0.
	\label{MEQ}
\end{split}
\end{align}

Thus, instead of employing Equations \eqref{EqK3} and \eqref{EqK4} to reduce the NU group to the Poincar\'e group in Minkowski spacetime, one can consider the equivalent differential constraints with Equation \eqref{MEQ}.

\subsubsection{General Case: Asymptotically Flat Spacetime with~Radiation}

As was mentioned above, the~NU group is an infinite-dimensional group, and therefore, there is no existing unique way to reduce the NU group to the finite-dimensional group even in Minkowski spacetime. All attempts suffer in supertranslational ambiguity. To~overcome this issue, we impose on $\xi_c$ the differential constraints restricting the NU group to a particular Poincar\'e quasigroup~\cite{24}:
\begin{align}\label{PEQ1} 
	\eth^2 \xi_s = ~& \frac{3}{2}\sigma^0\eth\xi_1 + \eth \sigma^0 \xi_1,  \\
	\bar \eth \xi_1 =~&0.
\label{PEQ2}
\end{align}

In our research, rather than using the complete infinite-dimensional BMS~\cite{57,58,59,60,61,62} or Weyl BMS group, we constrain the NU group to a finite-dimensional Poincaré quasigroup by imposing geometric constraints sourced from the geodesic deviation equation. A~principal advantage of the quasigroup formalism lies in its intrinsic capability to manage time-dependent structure functions inherent in radiating spacetimes. Concurrently, the~constraints enforced by Equations~(\ref{PEQ1}) and (\ref{PEQ2}) serve to remove supertranslation~ambiguities.

Since the spin weight of the asymptotic shear $\sigma^0$ is two, one can write $\sigma^0=\eth^2 \eta$, where $\eta = \eta(u,\zeta,\bar \zeta)$ is a complex function. Then, a general solution of Equation \eqref{PEQ1} can be written as
\begin{align}\label{K1a}
\xi_s=\xi_t + \eth\eta \xi_1 + \frac{u-\eta}{2}\eth\xi_1,
\end{align}
where $\eth^2\xi_t=0$.

We consider Equation~(\ref{PEQ1}) the differential constraint restricting the NU group to a particular Poincar\'e quasigroup~\cite{24}. In~the absence of radiation, the~differential constraint Equation \eqref{PEQ1} is compatible with the Killing equations, and~the Poincar\'e quasigroup becomes the Poincar\'e group. Note that the same constraint was obtained in~\cite{42,43} in the twistor theory~framework.

The structure of the Poincar\'e quasialgebra is as follows:

The generators of translations are given by
\begin{equation}
\xi_a = \xi_{t a}+ \bar \xi_{ta}=B_a\Delta^0,
\label{eqTr}
\end{equation}
where $B_a=\bar B_a$, with $B_a$ being the solution of the following equation:
\begin{align}
	 \eth^2 B_a =  0.
\end{align}

There are four independent solutions of this equation if~we assume that $\Im B_a$~=~0.

The generators of boosts and rotations are given by
\begin{align}
	\xi_A  = \xi_{cA} + \bar   \xi_{cA},
\end{align}
where
\begin{equation}
\xi_{cA} =(\eth\eta \bar C_A + \frac{u-\eta}{2}\eth \bar C_A)\Delta^0 + \bar C_A\delta^0, \label{eqBR}
\end{equation}
and $\eth C_A=0$. There are six independent solutions of the equation $\eth C_A=0$ if~we impose additional constraints:
\begin{align}
	&\bar \eth C_A + \eth \bar C_A=0 \quad (\rm rotations), \\
	&\bar \eth C_A - \eth \bar C_A=0  \quad (\rm boosts),
\end{align}

The straightforward computation shows that the generators of the Poincar\'e quasigroup obey at
commutation relations ${\mathscr{I}}^{+}$:
\begin{align}
& [ \xi_a ,\xi_b ]=0, \quad
[\xi_a,\xi_B]=C^b_{aB}(u,\zeta,\bar\zeta)\xi_b, \\
& \left [ \xi_A ,\xi_B \right]
= C^D_{AB}(u,\zeta,\bar\zeta)\xi_D,
\end{align}
where $C^D_{AB}$ and $C^b_{aB}$  are the structure functions. The~last commutation relations mean that the generators of the Lorentz quasigroup form a closed~algebra.

\section{Energy Momentum and Angular Momentum at \boldmath{$\mathscr{I}^{+}$}}\label{sec4}

As is well known, the~general covariance of the Einstein equations results in differential conservation laws related to the field equations. From~Noether's theorem, it follows that for an arbitrary diffeomorphism generated by the vector field $\xi$, the~invariance of the Lagrangian for the Einstein equations leads to the conservation laws of the form
\begin{equation}
{\frak J}^\mu(\xi)_{,\mu}=0,
\end{equation}
where the vector density ${\frak J}^\mu$ is defined as ${\frak J}^\mu(\xi)={\frak h}^{\mu\nu}_{,\nu}$, with~${\frak h}^{\mu\nu}=-{\frak h}^{\nu\mu}$ being the superpotential, which is constructed from the densities of the spin, bispin, vector field $\xi$,
and its derivatives~\cite{1,2,3}.

For the Einstein--Hilbert action, one obtains a simple expression of up to a factor of 2, \begin{equation}
{\frak J}^\mu= -\frac{1}{4\pi}\left(\sqrt{-g}\xi^{[\mu;\nu]} \right)_{,\nu}
\label{K} 
\end{equation} 
that was first given by Komar~\cite{4}. It is impossible to simply ``renormalize'' ${\frak J}^\mu$ by a factor of 2 because the normalization for energy momentum and angular momentum differs by a factor of 2. The~resolution of this problem is known as pointed out in~\cite{5}. One needs to add to the Einstein--Hilbert action $I$ a surface term $I_s$ and apply Noether's theorem to $I + I_s$. Komar's expression provides a fully satisfactory notion of the total mass in stationary, asymptotically flat~spacetimes. 

As is known, the~Komar integral is not invariant under a change in the choice of generators of time-like translations in the equivalence class associated with a given BMS translation. Moreover, the~resulting energy would not be the Bondi energy, but instead the Newman--Unti energy~\cite{36}. The~Bondi and Newman--Unti masses evaluated at null infinity are crucial for comprehending conserved quantities. The~Bondi mass is associated with the standard BMS time-like translation and shows a monotonic decrease. From the~other side, the~Newman--Unti mass uses an arbitrary generator that is neither unique nor necessarily monotonic, since there are infinitely many such functions (related by supertranslations).

Our approach ensures that $\mathcal{L}_{\xi}$ reduces to the Bondi mass linked with pure time translations, rigorously formulated within the Newman--Penrose framework, thereby ensuring adherence to pertinent flux balance laws. This is distinguished by its gauge independence attributable to the use of a null tetrad rather than coordinate systems. However, when addressing coordinate-dependent quantities such as the Bondi or NU mass, it is imperative to choose a specific gauge. The~constraints in Equations~(\ref{PEQ1}) and (\ref{PEQ2}) perform a role akin to establishing the Bondi gauge, identifying a specific ``radiation-adapted'' reference frame that facilitates the interpretation of conserved quantities. In~the absence of these constraints, computation of the Komar-type integral (Equation (\ref{eqn: Equation(101)})) using an arbitrary BMS generator would yield a quantity similar to the NU energy, accompanied by its intrinsic gauge~ambiguities.

According to the framework developed by Nester~et~al.~\cite{63}, the~gravitational energy momentum is linked to the Hamiltonian by applying specific boundary conditions. Pitts~\cite{65} showed that utilizing Noether's first theorem leads to an infinite set of conserved~currents. 
         
Therefore, we do not claim to identify the definitive energy and angular momentum among the infinitely many alternatives. Instead, we propose a geometric criterion that is particularly suitable for the analysis of radiating isolated systems at future null infinity and for resolving the supertranslation ambiguity in angular~momentum.
        
Our results should be situated within the contemporary framework of gravitational energy momentum in terms of the function $\lambda^{0}$, which represents information emerging from the positive null infinity $\mathscr{I}^{+}$ and $\sigma^{0}$ corresponds to the shear-free parameter. With~shear $h=h_{+}-ih_{\times}\propto\sigma^{0}$,~the news function is linked to its time-derivative $\dot{h}\propto\lambda^{0}$ gravitational wave polarizations $h$ from gauge-invariant perturbations derived by Thorne and Ruiz~\cite{52} and Thorne~\cite{25}. 

For an asymptotically flat at future null infinity spacetime, the modified ``gauge-invariant'' 
 Komar integral (linkage) was introduced by Tamburino and Winicour~\cite{6,7}. The~computation leads to the following coordinate-independent expression~\cite{11,12,13}:
\begin{align}
\mathcal  L_{\xi}&=-\dfrac{1}{4\pi}\Re
\oint [B\left(\Psi^0_2 + \sigma^0\lambda^0 -\eth^2\bar\sigma^0\right)
\nonumber\\
&+ \bar C\left(\Psi^0_1 -
\sigma^0\eth^2\bar\sigma^0 -(1/2)\eth(\sigma^0\bar\sigma^0)\right)]
d\Omega,
\label{eqn: Equation(101)}
\end{align}
where $\lambda^0 =\dot {\bar\sigma}^0 $, and~$\xi$ is an arbitrary generator of the NU group at $\mathscr{I}^+$,
\begin{align}
\xi = B \Delta^0 	+ \bar C \delta^0 +  C \bar \delta^0, \quad \eth C =0.
\end{align}
        
A complex generator,  $\xi_c =\xi_s\Delta^0 + \xi_1 \delta^0 $, of~the Poincar\'e quasigroup yields the complex Noether charge that can be written as~\cite{10}
\begin{align}
Q_{\xi_c}&=-\frac{1}{8\pi} \oint\{\xi_s\left(\Psi^0_2 - \sigma^0\lambda^0
-\eth^2\bar\sigma^0\right)\nonumber\\
&+ \xi_1\left(\Psi^0_1 - \sigma^0\eth{\bar\sigma}^0
-(1/2)\eth(\sigma^0\bar\sigma^0)\right)\} d \Omega .
\label{NQ}
\end{align}

The complex Noether charge $Q_{\xi_{c}}$ encodes both the real conserved quantity and phase information. The~physical (real) conserved quantity is defined by $\mathcal L_\xi$. 

Setting $B =(\xi_s + \bar\xi_s)/2$ and $ \bar C_1 = \xi_1$, we obtain
\begin{equation}
\mathcal L_\xi= Q_{\xi_c}+  \overline Q_{\xi_c} =  Q_{\xi_c}+  Q_{\bar\xi_c}.
\end{equation}

The factor $1/8\pi$ in Equation \eqref{NQ} is determined by calculating the intrinsic angular momentum, $J=ma$, for~the Kerr metric with the rotational Killing vector, $L_z $, yielding $\xi_1= - L_z  \cdot \bar m ={i}/{\sqrt{2}} \sin \theta$. The~Weyl scalar, $\psi^0_1 = \sigma/\sqrt{2}$ is computed with respect to a Bondi frame with shear-free cross-sections~\cite{10, 44}.

As in the previous section, to~reduce the NU group to the Poincar\'e quasigroup, we impose the following differential constraints:
\begin{align} \label{PEQ4a}
	\eth^2 \xi_s = ~& \frac{3}{2}\sigma^0\eth\xi_1 + \eth \sigma^0 \xi_1, \\
	\bar \eth \xi_1 =~&0.
\label{PEQ4b}
\end{align}

One can show that the following integral identity is valid:
\begin{align}
	\oint \eth^2 \xi_s \bar \sigma^0d \Omega = 	\oint \xi_1 \Big(\eth \sigma^0\bar \sigma^0  - \frac{3}{2}\eth(\sigma^0 \bar \sigma^0 )\Big )d \Omega .
\end{align}

Using this identity in Equation \eqref{NQ}, one can rewrite it in the equivalent form introduced in~\cite{43}:
\begin{align}
Q_{\xi_c}&=-\frac{1}{8\pi} \oint\{\xi_s\left(\Psi^0_2 + \sigma^0\lambda^0
+\eth^2\bar\sigma^0\right)\nonumber\\
&+ \xi_1\left(\Psi^0_1 + \sigma^0\eth{\bar\sigma}^0
+(1/2)\eth(\sigma^0\bar\sigma^0)\right)\} d \Omega .
\label{NQ1}
\end{align}

Now, employing Equations \eqref{PEQ4a} and \eqref{PEQ4b} and performing integration by parts, we find that Equation \eqref{NQ1} can be recast as
\begin{align}
Q_{\xi_c} =-\frac{1}{8\pi} \oint\Big(\xi_s (\Psi^0_2 + \sigma^0\lambda^0 )+ \xi_1\Psi^0_1 \Big ) d \Omega.
\label{NQ2}
\end{align}

We adopt this as the definition of the conserved quantities on $\mathscr{I}^+$ associated with the generators of the Poincar\'e quasigroup, writing~\cite{10,24,43,45}
\begin{align}
	 \mathcal  L_\xi= Q_{\xi_c}+  \overline Q_{\xi_c} .
\end{align}

The integral four-momentum is given by
\begin{equation}
    P_a=  -\frac{1}{4\pi} \Re \oint l_a (\Psi^0_2 + \sigma^0\lambda^0 ) d\Omega.
    \label{EM}
\end{equation}
where $l^{a}$ denotes the four independent solutions of $\eth^{2}l^{a}=0$ representing pure translations, $\Psi^{1}_{2}$ is the leading-order Weyl scalar (mass aspect), $\sigma^{0}$ is the asymptotic shear (encodes gravitational radiation), $\lambda^{0}=\dot{\overline{\sigma}}^{0}$ is the news functions (rate of change of shear), and~some of the physical components are denoted by $P^{0}=M$ (Bondi mass). 
 The total energy/mass of the system and $P^{1}$,~$P^{2}$, and~$P^{3}$ are the components of linear momentum in three spatial~directions. 

Using the Bianchi identities, we compute the loss of energy momentum and obtain the standard expression for the flux balance law (see, e.g., \cite{13}):
\begin{equation}
            \dot P_a = -\left(\dfrac{1}{4\pi}\right) \oint l_a |\lambda^0|^2 d\Omega.
            \label{EML}
        \end{equation}
        
This shows that energy/momentum decreases monotonically during gravitational wave emission.

The angular momentum is given by
\begin{equation}
           M_i=-\frac{1}{4\pi} \Re \oint\Big(\xi_i (\Psi^0_2 + \sigma^0\lambda^0 )+ \bar L_i\Psi^0_1 \Big )d \Omega ,
        \label{AM}
        \end{equation}
where $L_{i}$ denotes the three rotation generators on $S^{2}$, $\Psi^{1}_{0}$ is the Weyl scalar component related to angular momentum, and $\xi_i$ is the solution of Equation~(\ref{PEQ4a}) (adapted to radiation), 
\begin{align}
\xi_{i}= \eth\eta \bar L_i + \frac{u-\eta}{2}\eth\ \bar L_i .
\label{GR}
\end{align}

Here, $L_i$ is the solution of the equation $\eth L_i = 0$ such that $\bar \eth  L_i  + \eth \bar L_i  =0$ and~$\sigma^0 = \eth^2 \eta$.

Substituting $\xi_i$ in Equation \eqref{AM} and performing integration by parts, we find
\begin{equation}
M_i=-\frac{1}{4\pi} \Re \oint \bar L_i{\cal J} dS \Omega,
\label{AM1}
\end{equation}
where
\begin{align}
 {\mathcal J}=~&\Psi^0_1  +\frac{3}{2}\eth\eta(\Psi^0_2 + \sigma^0\lambda^0 )+\frac{\eta}{2}\big(\eth\Psi^0_2 + \eth(\sigma^0\lambda^0)\big ).
\end{align}

It yields the following expression for the angular momentum loss:
\begin{align}
\dot M_i = -\frac{1}{4\pi} \Re \oint \bar L_i
\frac{\partial {\cal J}}{\partial u} d\Omega .
\end{align}

Geroch and Winicour have given a list of properties that conserved quantities $P(\xi, \Sigma)$ defined at $\mathscr{I}^{+}$ and should have the following~\cite{11}: 
\begin{itemize}

\item $P(\xi, \Sigma)$ should be linear in the generators of the asymptotic symmetry~group.
            
\item $P(\xi, \Sigma)$ should be invariant with respect to Weyl rescalings of the form $\tilde{g}_{\mu\nu}=\tilde{\Omega}^{2}g_{\mu\nu}$, where $\tilde{\Omega}$ constitutes a smooth function that is non-vanishing across the conformal~transformation.

\item  The expression $P(\xi,\Sigma)$ should depend on the geometry of
$\mathscr{I}^+$  and behavior of generators in the neighborhood of $\mathscr{I}^+$.

\item  $P(\xi,\Sigma)$ should be proportional to the corresponding Komar integral for the
exact symmetries and coincide with the Bondi four-momentum when $\xi$ is a
BMS~translation. 

\item  $P(\xi,\Sigma)$ should be defined also for the system with radiation on $\mathscr{I}^+$.

\item  There should exist a flux integral $\cal I$ that is linear in
$\xi$ and that gives the difference $P(\xi,\Sigma') - P(\xi,\Sigma)$ for~$\Sigma'$ and $\Sigma'$, two closed surfaces on $\mathscr{I}^+$. 

\item In Minkowski spacetime, $P(\xi,\Sigma)$ should vanish identically.
\end{itemize}

Our definition of the conserved quantities, $\mathcal  L_\xi= Q_{\xi_c}+  \overline Q_{\xi_c} $, where the complex Noether charge is given by Equation \eqref{NQ2}, is free from the supertranslation ambiguity and satisfies all these~conditions.

\section{Center of Mass and Intrinsic Angular~Momentum}
\label{COM}

In special relativity, total angular momentum is given by $J^{\mu \nu} = X^{\mu }P^{\nu} - 
P^{\mu }X^{\nu}  + S^{\mu \nu}$, where the two first terms are the angular momentum, and~$ S^{\mu 
\nu}$ denotes intrinsic angular momentum, with~$X^\mu$ and $P^\nu$ being the position 
vector and momentum of the particle, respectively. Using space-like translation freedom, one 
can transform the given reference frame to the center-of-mass reference frame (CMRF), where 
$P^\nu$ is aligned with the observer's four-velocity. In~the CMRF, one has $J^{\mu \nu} 
P_\nu=0$. Applying the transformation $X^\mu = P^\mu/P_\nu P^\nu$, we obtain the Dixon 
condition on the intrinsic angular momentum, $S^{\mu \nu} P_\nu=0$ \cite{46}. 

The Dixon condition implies that the boost generators do not contribute to the internal part of the total angular momentum~\cite{26}. At~future null infinity, this condition reduces to the requirement that the linkage $L_\xi =0$ for the boost generators. We adopt this requirement for the general case by imposing the condition that the Noether charge $K_i= 0$ for the boost generators:
\begin{align}
K_{i} =-\frac{1}{4\pi} \Re\oint\Big(\xi^R_i (\Psi^0_2 + \sigma^0\lambda^0 )+ \bar R_i\Psi^0_1 \Big ) d \Omega,
\label{NQ2a}
\end{align}
where
\begin{align}
\xi^R_{i}= \eth\eta \bar R_i + \frac{u-\eta}{2}\eth \bar R_i ,
\end{align}
and $R_i$ is the solution of the equation $\eth R_i = 0$ such that $\bar \eth  R_i  - \eth \bar R_i  =0$.

As one can see,~quantities  $R_i$  and $L_i$ are related by relations $R_i = \pm i L_i$ (see Table~\ref{tab:placeholder} in Section~\ref{sec3}). Then, the condition $K_i =0$, or~Dixon condition, can be written as
\begin{align}
            \Im\oint\Big(\xi_i (\Psi^0_2 + \sigma^0\lambda^0 )+ \bar L_i\Psi^0_1 \Big ) d \Omega =0,
            \label{NQ3a}
        \end{align}
where $\xi_i$ is defined by Equation \eqref{GR}. In~terms of the complex Noether charge,  \mbox{Equation \eqref{NQ3a}} can be recast as $\Im Q_{\xi_c} =0$. This condition defines the center-of-mass frame at null infinity by requiring that boost charges vanish. In~the framework of special relativity, the~center-of-mass reference frame is characterized by the vanishing of total spatial momentum: $\overline{P}=0$. Our condition represents a generalization applicable to asymptotically flat spacetimes in the presence of gravitational~radiation.

To transform the given reference frame at null infinity to the CMRF frame, we use supertranslation freedom to align the total momentum with the observer's four-velocity. To~explain in detail, let us consider the Noether charge associated with the mass aspect,
\begin{align}
\mathcal Q_a = \oint\xi_a\Psi d \Omega  =\oint\xi_a (\Psi^0_2 + \sigma^0\lambda^0 + \eth^2 \bar \sigma^0) d \Omega .
\label{NQ3}
\end{align}

In the CMRF, only component $\mathcal Q_0 $ is different from zero. Thus, to~align the total momentum with the observer's four-velocity,  one should require $\mathcal Q_i =0$ for $i=1,2,3$. Using this condition in Equation~\eqref{NQ3a}, we obtain
\begin{align}
\Im\oint \big( \bar L_i\Psi^0_1  -\xi_i  \eth^2 \bar \sigma^0\big ) d \Omega =0,
\label{NQ3b}
\end{align}
where the integration is performed over the section defined by equation $u- \eta_{0}=0$.  
 Further simplification can be performed by employing Equation \eqref{PEQ4a}. The~computation yields
\begin{align}
\Im\oint\bar L_i\big(2\Psi^0_1  -\eth\sigma^0 \bar \sigma^0 - 3 \sigma^0\eth \bar \sigma^0 \big ) d \Omega =0.
\label{CM1}
\end{align}

The obtained results are an integral expression of the condition defining the center of mass deduced in~\cite{17}. This corresponds to the Dixon condition imposed on the intrinsic angular momentum, $S^{\mu \nu} P_\nu=0$.

To proceed further, we expand $\xi_i$, $\sigma^0$, and $\eta$ in terms of spin-weighted spherical harmonics:
\begin{align}
\xi_i=~&\sum^\infty_{l=0} \sum^l_{m=-l}  \xi^i_{lm} Y_{l m}
	\label{eta1a}, \\
\sigma^0=~& \sum^\infty_{l=2} \sum^l_{m=-l}  h_{lm} (u) \, {}_2Y_{l m}, \label{IAC1a}\\
		\eta=~&\eta_0 +  \sum^\infty_{l=1} \sum^l_{m=-l} \eta_{lm}Y_{lm}.
		\label{IAC1b}
\end{align}

From the relation $\sigma^0 = \eth^2 \eta$, it follows that
\begin{align}
	 \eta_{lm} = \frac{2 h_{lm} }{\sqrt{l(l^2-1)(l+2)} }, \quad  l \geq 2 .
	 \label{eta}
\end{align}
	 
We use the freedom in the choice of $\eta$ to eliminate in Equation \eqref{eta1a} the contribution of the term with $l=0$, choosing the cut as $u- \eta_0 =0$. We obtain
\begin{align}
\xi_i= &\sum^\infty_{l=1} \sum^l_{m=-l}  \xi^i_{lm} Y_{l m}
	\label{eta1}.
	\end{align}
 Using Equation \eqref{eta1} in Equation \eqref{NQ3}, we have
\begin{align}
\mathcal J_i = \sum^\infty_{l=1} \sum^l_{m=-l}  \xi^i_{lm} \oint Y_{lm} (\Psi^0_2 + \sigma^0\lambda^0 + \eth^2 \bar \sigma^0) d \Omega .
\end{align}

To proceed further, we consider  the supermomenta $P_{lm} (\Sigma)$ introduced in~\cite{20},
\begin{align}
P_{lm} (\Sigma)= -\frac{1}{\sqrt{4\pi}}	\int_\Sigma Y_{lm} \Psi d \Omega,
\end{align}
where $\Sigma$ is an  arbitrary section of $\mathscr{I}^+$. A~nice section is defined by the requirement that the supermomentum $P_{lm} (\Sigma)=0$ for $l \geq 1$.  Thus, only the component $P_0$ of the total momentum is non-vanishing, providing us with a geometric notion of a reference frame at rest. This, in~general, involves the need to make a Lorentz boost, which keeps $\Sigma$ fixed and aligns the generator of time translations with the total momentum~\cite{21}. As~one can see, for~the nice sections, the quantity $\mathcal J_i =0$. 

Employing the condition $\mathcal J_i =0$ in Equation \eqref{AM}, after~some computations, we find that the intrinsic angular momentum, $J_i$,  can be written as the Komar angular momentum:
\begin{equation}
            J_i=-\frac{1}{4\pi} \Re \oint \bar L_i\big( \Psi^0_1 + \sigma^0\eth \bar \sigma^0 \big )d \Omega .
            \label{KAM}
        \end{equation}  
              
This is the angular momentum in the center-of-mass frame where orbital angular momentum vanishes. An~independent calculation by Gallo and Moreschi~\cite{21} confirms Equation~\eqref{KAM}.

Hence, the angular momentum loss is given by
\begin{equation}
\dot J_i=-\frac{1}{4\pi} \Re \oint \bar L_i\big( \dot \Psi^0_1 + \bar \lambda^0\eth \bar \sigma^0 + \sigma^0\eth {\lambda}^0\big )d \Omega ,
\label{IAM}
\end{equation}

Using the Bianci identities~\cite{30,34},
\begin{align}
& \Psi^0_2 -\bar\Psi^0_2 = \bar\sigma^0\bar\lambda^0 - \sigma^0\lambda^0
+ \bar\eth^2\sigma^0 - \eth^2\bar\sigma^0, \label{Psi2} \\
&\dot\Psi^0_1 = \eth \Psi^0_2 - 2 \sigma^0\eth\lambda^0,
\end{align}

Equation \eqref{IAM} reduces to
\begin{align}
\dot J_i&=-\frac{1}{8\pi} \Re \oint \bar L_i\big(  3\bar \lambda^0\eth \bar \sigma^0 -3 \sigma^0\eth {\lambda}^0
\nonumber\\
&+  \bar\sigma ^0\eth {\bar \lambda }^0  - \lambda^0\eth {\sigma}^0 \big )d \Omega .
\label{IAM1}
\end{align}

This expression agrees with the results obtained in~\cite{47}.

To compute the angular momentum loss, we express $\sigma^0$ and $\bar L_i$ in terms of spin-weighted spherical harmonics:
\begin{align}
	\sigma^0=& \sum^\infty_{l=2} \sum^l_{m=-l}  h_{lm} (u)\, {}_2Y_{l m},\\
	\bar L_i=&  \sum^1_{m=-1}  L_{im} \, {}_{-1}Y_{l1m}.
	\label{eta7}
\end{align}

The computation yields (see Appendix \ref{sec: Appendix B} for details)
\begin{align}
	\dot J_i = ~&\frac{1}{8\pi}  \Re\sum_{l,m,l',m'} \big ( h_{lm} \dot{\bar h}_{l'm'} -   \dot h_{lm} {\bar h}_{l'm'} \big) c^i_{lml'm'} \nonumber \\
	=~&\frac{1}{8\pi}  \Re\sum_{l,m,l',m'} h_{lm} \dot{\bar h}_{l'm'} \big (c^i_{lml'm'} - \bar{c^i}_{l'm'lm}   \big) ,
	\label{ACML}
\end{align}
where the coefficients $c^i_{lml'm'}$ are given in terms of the Wigner 3-$j$ symbols: 
\begin{widetext}
\begin{align}
c^i_{lml'm'} =
&(-1)^{m' }\sqrt{\frac{3(2 l+1)(2 l'+1)}{8 \pi}}
\bigg\{
3\sqrt{(l'+2)(l'-1)}
\left(\begin{array}{ccc}
l & l' & 1\\
-2 & 1 & 1
\end{array}\right) \nonumber \\
 & +\sqrt{(l+3)(l-2)}
\left(\begin{array}{ccc}
l & l' & 1\\
-3 & 2 & 1
\end{array}\right) \bigg\}
\sum^1_{m''=-1}L_{im''} \left(\begin{array}{ccc}
l & l' & 1\\
m & -m'& m''
\end{array}
\right).
\end{align}	
\end{widetext}

In Appendix \ref{sec: Appendix B}, it is demonstrated that, for~the specific case $l=2$, the~loss of the $z$-component of angular momentum is represented by the equation
\begin{align}
	   \dot J_z =~&\frac{1}{4\pi}  \Im\sum^{m=2}_{m = -2}m h_{2m} \dot{\bar h}_{2m}.
	   \label{BJz}
\end{align}
where $h_{lm}$ denotes the spherical harmonic decomposition coefficients of the~shear.

Finally, in~the last equation, we obtain the momenta carried away by the terms of the gravitational wave polarizations {$h$} from gauge-invariant perturbations derived by Thorne in~\cite{47} and Lousto~\cite{25}.

The angular momentum, as~expressed in Equation~(\ref{BJz}), represents a classical conserved quantity that has been extensively studied in relation to various astrophysical entities, including black holes, accretion disks, the~collapse of black holes, supernovae, and~gravastars, among~others. In~the context of quantum gravity applications, the~concept of gravitational memory $\sigma^0$ emerges as a compelling subject for investigation, particularly concerning gravitational wave bursts with memory, higher-dimensional frameworks,~nonlinear characteristics of gravitation, and~gravitational wave experiments.
Our goal in this paper was to demonstrate that the Poincaré quasigroup algebra in a different way to obtain the classical angular moment and that allows us to obtain the energy momentum, angular momentum, and~gravitational memory, which can be used for several physical examples in gravitational waves and other topics. These quantities have been demonstrated using various formalisms. For~instance, Milton~et~al.~\cite{6} employed the decomposition of the Weyl scalar $ \psi_4 $ into spin-weighted spherical harmonics. Similarly, Lousto~et~al.~\cite{45} derived the outgoing radiation, represented by the Weyl scalar $ \psi_4 $ within the Kinnersley tetrad, as~discussed by Thorne, K.~\cite{46} in terms of gauge-invariant perturbations.
In \mbox{Appendix \ref{apendixA}}, we present a derivation of the spin-weighted spherical harmonics, and \mbox{Appendix \ref{sec: Appendix B}} addresses the angular momentum analysis within the context of the Poincaré quasigroup, drawing upon the different wave treatment of Milton, Lousto, and~Thorne.
This observation presents an excellent opportunity for further exploration in future studies utilizing our formalism, but~nevertheless, the main focus of this research was to attain angular momentum in a nonlinear~representation.

\section{Discussion and~Conclusions}\label{sec6}

In this paper, we address the task of identifying conserved quantities, such as energy momentum, angular momentum, and linear moment radiation with an isolated system, within~asymptotically flat spacetimes as governed by general relativity. 
 Due to the complex nature of gravitational radiation, gauge choices, or~field-dependent transformations being too restrictive with traditional group structures,~quasigroup use offers a better, more transparent way to describe the algebra of asymptotic symmetries. 
  Our investigation demonstrates that the application of quasigroup symmetries facilitates a systematic and unified representation of the conserved~quantities.

By employing the quasigroup methodology, we developed a framework that effectively resolves the conventional challenges presented by supertranslation ambiguity at null infinity. Our method facilitates the establishment of a geometric and algebraic structure invariant under supertranslations; in particular, we derive the intrinsic angular momentum and delineate it with the center-of-mass frame reference. In~conjunction with methodologies for modulating supertranslation freedom, this highlights the pivotal importance of selecting suitable sections of null infinity for rigorous physical~analysis.

Utilizing methodologies such as the spin coefficient formalism and Newman--Penrose scalar quantities, in~conjunction with an in-depth analysis of the Bondi--Metzner--Sachs group, we derived explicit formulae that describe the loss of angular momentum and energy conveyed by gravitational radiation with the Weyl tensor $\Psi_4$ in terms of the spin-weighted spherical harmonics. It is important to remark that our results for conservative quantities are well known and in concordance with established results using other~methodologies.

Furthermore, by~employing Poincaré quasigroups, we identified conserved quantities in terms of polarizations $h_{+}$ and $h_{\times}$, which are directly related to the physics of gravitational waves in astronomy, specifically for estimating parameters of black holes, neutron stars, and~supernovae. In~future work, these quantities could be obtained using data gathered from the detectors of the LIGO/Virgo/KAGRA international collaboration. Using tools such as numerical simulations in relativity allows for standardizing waveform extraction methods;
in the field of black hole physics, for~verifying the Kerr no-hair theorem and quantifying spin parameters; and in~the domain of quantum gravity, for interconnections with BMS symmetries, soft graviton theorems, and~the black hole information~paradox. 

This investigation enhances our understanding of the asymptotic structure of spacetime and offers a dependable methodology for distinctly identifying conserved quantities within gravitational systems. Subsequent research endeavors might extend these concepts to encompass more generalized frameworks, such as spacetimes affected by gravitational forces and intricate matter configurations, thereby augmenting both theoretical insights and their astrophysical~significance.

\acknowledgements

A.Z.R. acknowledges a SECIHTI scholarship. C.M. and A.I.N. want to thank SNII-SECIHTI, PROINPEP-UDG, and PROSNII-UDG. 

\appendix

\section{Spin-Weighted Spherical~Harmonics}\label{apendixA}

A quantity $\eta$ is said to have a spin weight $s$ (s.w. $s$) if~$\eta$
transforms as $\eta \rightarrow e^{is\theta}\eta$ under a rotation of
$m^\alpha \rightarrow e^{i\theta}m^\alpha$, with~$\theta$ being a real function
of position. The~complex conjugate $\bar \eta$ has spin weight $-s$. The~spin coefficients of the set $\{\rho,\mu,\epsilon + \bar\epsilon, \gamma +
\bar\gamma\}$   have an s.w. of zero, elements of $\{\kappa,\tau,\bar\alpha +
\beta,\nu,\bar\pi\}$ have an s.w. of $+1$, and elements of
$\{\sigma,\bar\lambda\}$ have an s.w. of $+2$ \cite{9}.

For a function $\eta(u,\zeta,\bar\zeta)$ of spin weight $s$, the~raising and  lowering  operators  $\eth$ and $\bar\eth$, respectively,  are defined on  $\mathscr{I}^+$ as follows~\cite{17,32}:
\begin{align}
\begin{split}
\eth \eta : = \delta^0 f + 2\bar\alpha^0 s f = P{\bar
P}^{-s}\frac{\partial (\bar P^s  \eta )}{\partial \zeta},\nonumber \\
\bar\eth  \eta : = \bar\delta^0 f - 2\alpha^0 s f = \bar P{P}^{s}\frac{\partial
(P^{-s}  \eta )}{\partial \bar\zeta},
\label{eth}	
\end{split}
\end{align}
where
\begin{align}
	\delta^0 = P\frac{\partial}{\partial \zeta } , \quad
	\alpha^0 =\frac{P}{2}\frac{\partial \ln \bar P}{\partial \zeta } .
\end{align}

Here, $P$ is a conformal factor emerging from a two-dimensional
metric of cuts on $\mathscr{I}^+$: 
\begin{align}
	ds^2 = \frac{2d\zeta d\bar\zeta}{|P(u,\zeta,\bar\zeta)|^2}.
\end{align}

In a particular choice of $P=P_0=(1+\zeta\bar\zeta)/\sqrt{2}$, the metric
above will be the metric of the two-sphere, and~the coordinates are referred to as
Bondi coordinates (BCs). 

 The commutator $[\eth,\bar \eth]$ is given by
\begin{align}
	[\eth,\bar \eth]\eta =
s\eta\eth\bar\eth\ln(P\bar P)=(K + \bar K)s\eta,
\end{align}
where $(K + \bar K)$ is the Gaussian curvature of a two-dimensional cut. The~following useful lemma holds:

\begin{Lemma}
If 
 $\text{\rm s.w. }A+\text{\rm s.w. }B = -1$, then
\begin{align}
	\oint A\eth B d \Omega = -\oint B\eth A d \Omega.
\end{align}
where s.w. denotes spin weight~\cite{10}.
\end{Lemma}

In the following, we make a choice of the conformal factor as $P = P_0 \sqrt{\zeta/\bar \zeta}$. This yields
\begin{align}
\delta^0=\frac{1 +\zeta \bar \zeta}{\sqrt{2}}  \sqrt{\frac{\zeta}{\bar \zeta}} \;\frac{\partial}{\partial \zeta}.
\end{align}

In the spherical coordinates, we obtain
\begin{align}
\delta^0=-\frac{1}{\sqrt{2}}\left(\frac{\partial}{\partial \theta} +  \frac{i}{\sin \theta}\frac{\partial}{\partial \varphi}\right).
\end{align}

\begin{widetext}
This gives, up~to the factor  $1/\sqrt{2}$, the~standard form of raising and lowering operators~\mbox{\cite{32,33,50}}:
\begin{align}
\eth \eta=-\frac{1}{\sqrt{2}}\left(\partial_{\theta}+\frac{\mathrm{i}}{\sin \theta} \partial_{\phi}-s \cot \theta\right) \eta=-\frac{1}{\sqrt{2}}\sin ^{s} \theta\left(\partial_{\theta}+\frac{\mathrm{i}}{\sin \theta} \partial_{\phi}\right)\left(\eta \sin ^{-s} \theta\right), \\
\bar{\eth} \eta=-\frac{1}{\sqrt{2}}\left(\partial_{\theta}-\frac{\mathrm{i}}{\sin \theta} \partial_{\phi}+s \cot \theta\right) \eta=-\frac{1}{\sqrt{2}}\sin ^{-s} \theta\left(\partial_{\theta}-\frac{\mathrm{i}}{\sin \theta} \partial_{\phi}\right)\left(\eta \sin ^{s} \theta\right),
\end{align}

A direct computation shows that
\begin{align}
&\bar \eth \eth\eta=\frac{1}{2}\left(\frac{1}{\sin \theta} \partial_{\theta} \sin \theta \partial_{\theta}+\frac{1}{\sin ^{2} \theta} \partial_{\phi}^{2}+\frac{2 \mathrm{is} \cos \theta}{\sin ^{2} \theta} \partial_{\phi}-\frac{s^{2}}{\sin ^{2} \theta}+s(s+1)\right) \eta,\\
&\eth\bar \eth\eta=\frac{1}{2}\left(\frac{1}{\sin \theta} \partial_{\theta} \sin \theta \partial_{\theta}+\frac{1}{\sin ^{2} \theta} \partial_{\phi}^{2}+\frac{2 \mathrm{is} \cos \theta}{\sin ^{2} \theta} \partial_{\phi}-\frac{s^{2}}{\sin ^{2} \theta}+s(s-1)\right) \eta,
\end{align}

Let ${}_sY_{j m}$ be a normalized spin-weighted spherical harmonic
of order $j$ and spin weight $s$; then, we have,  $s \bar{Y}_{l m}=(-1)^{s+m} {}_{-s} Y_{l -m}$, and~\begin{align}
\eth\, {}_sY_{j m}=&\sqrt{  \frac{j(j+1)-s(s+1) }{2} }  \,\,{}_{s+1}Y_{j m} ,\\
\bar{\eth} \,{}_sY_{j m}=&-\sqrt{ \frac{j(j+1)-s(s-1)}{2}}  \,\,{}_{s-1} Y_{j m}.
\end{align}

When $s_1 + s_2 + s_3 = 0$, the triple integral is given in terms of the Wigner 3-$j$ symbols:
\begin{align}	
\int_{S^{2}}&{ }_{s_{1}} Y_{j_{1} m_{1} s_{2}} Y_{j_{2} m_{2} s_{3}} Y_{j_{3} m_{3}} d\Omega\nonumber \\
&=\sqrt{\frac{\left(2 j_{1}+1\right)\left(2 j_{2}+1\right)\left(2 j_{3}+1\right)}{4 \pi}}\left(\begin{array}{ccc}
j_{1} & j_{2} & j_{3} \\
m_{1} & m_{2} & m_{3}
\end{array}\right)\left(\begin{array}{ccc}
j_{1} & j_{2} & j_{3} \\
-s_{1} & -s_{2} & -s_{3}
\end{array}\right)
\label{ITr}
\end{align}
For the references, we present analytic expressions for the first few orthonormalized spin-weighted spherical harmonics~\cite{51}:
\end{widetext}

\begin{align}
{ }_{-1} Y_{10} &=-\sqrt{\frac{3}{8 \pi}} \sin \theta \\
{ }_{-1} Y_{1 \pm 1} &=-\sqrt{\frac{3}{16 \pi}}(1 \pm \cos \theta) e^{\pm i \varphi} \\
{}_{-2}Y_{20} &=\sqrt{\frac{15}{32 \pi}} \sin ^{2} \theta \\
{}_{-2} Y_{2 \pm 1} &=\sqrt{\frac{5}{16 \pi}} \sin \theta(1+\cos \theta) e^{ \pm i \varphi} \\
{}_{-2} Y_{2 \pm 2} &=\sqrt{\frac{5}{64 \pi}}(1+\cos \theta)^{2} e^{ \pm 2 i \varphi} 
\end{align}

Using the relation $s \bar{Y}_{l m}=(-1)^{s+m} {}_{-s} Y_{l -m}$, we obtain
\begin{align}
{ }_{1} Y_{10} &=\sqrt{\frac{3}{8 \pi}} \sin \theta \\
{ }_{1} Y_{1 \pm 1} &=-\sqrt{\frac{3}{16 \pi}}(1 \mp \cos \theta) e^{\pm i \varphi} \\
{}{}_{2}Y_{20} &=\sqrt{\frac{15}{32 \pi}} \sin ^{2} \theta \\
{}_{2} Y_{2 \pm1} &=-\sqrt{\frac{5}{16 \pi}} \sin \theta (1\mp\cos\theta) e^{ \pm i \varphi} \\
{}_{2} Y_{2 \pm 2} &=\sqrt{\frac{5}{64 \pi}}(1\mp\cos \theta )^{2} e^{\pm 2 i \varphi}
\end{align}

\begin{widetext}
\section{Angular Momentum~Loss}\label{sec: Appendix B}

The loss of the intrinsic angular momentum is given by
\begin{align}
\dot J_i=-\frac{1}{8\pi} \Re \oint \bar L_i\big(  3\bar \lambda^0\eth \bar \sigma^0 -3 \sigma^0\eth {\lambda}^0
+\bar\sigma ^0\eth {\bar \lambda }^0  - \lambda^0\eth {\sigma}^0 \big )d \Omega .
\label{IAC1}
\end{align}

To proceed further, we express $\sigma^0$ and $\bar L_i$ in terms of spin-weighted spherical harmonics:
\begin{align}\label{IAC1}
	\sigma^0=& \sum^\infty_{l=2} \sum^l_{m=-l}  h_{lm} \, {}_2Y_{l m},\\
		\bar L_i=&  \sum^1_{m=-1}  L_{im} \, {}_{-1}Y_{1m}.
\end{align}

Taking the derivative, we obtain
\begin{align}
\eth\sigma^0=& \sum^\infty_{l=2} \sum^l_{m=-l}  h_{lm} \sqrt{ \frac{(l-2)(l+3) }{2}}\, {}_3Y_{l m},\\
	\eth\bar\sigma^0=& -\sum^\infty_{l=2} \sum^l_{m=-l} \bar h_{lm} \sqrt{\frac{ (l+2)(l-1)}{2} }\, {}_1\bar Y_{l m}.
	\label{AC2}
	\end{align}
	
The computation yields
\begin{align}
	\dot J_i =  \frac{1}{8\pi}  \Re\sum_{l,m,l',m'} \big ( h_{lm} \dot{\bar h}_{l'm'} -   \dot h_{lm} {\bar h}_{l'm'} \big) c^i_{lml'm'} 
	= \frac{1}{4\pi}  \Re\sum_{l,m,l',m'} h_{lm} \dot{\bar h}_{l'm'} \big (c^i_{lml'm'} - \bar{c^i}_{l'm'lm}   \big) ,
\end{align}
where 
\begin{equation}
    \begin{split}
        &c^{i}_{lml'm'}=\\
    &\sum^1_{m''=-1}L_{im''}\oint d \Omega \ _{-1}Y_{1m''} \left(\sqrt{\frac{ (l+3)(l-2) }{2}}\ _{3}Y_{l m}\ _{2}\bar Y_{l' m'} - 3\sqrt{ \frac{(l'+2)(l'-1)}{2}}\ _{2}Y_{l m} \ _{1}\bar Y_{l' m'}\right),
    \end{split}
\end{equation}
and performing the integration, we obtain
\begin{align}
c^i_{lml'm'} =	&(-1)^{m' }\sqrt{\frac{3(2 l+1)(2 l'+1)}{8 \pi}}
\bigg\{
3\sqrt{(l'+2)(l'-1)}
\left(\begin{array}{ccc}
l & l' & 1\\
-2 & 1 & 1
\end{array}\right) \nonumber \\
 & +\sqrt{(l+3)(l-2)}
\left(\begin{array}{ccc}
l & l' & 1\\
-3 & 2 & 1
\end{array}\right) \bigg\}
\sum^1_{m''=-1}L_{im''} \left(\begin{array}{ccc}
l & l' & 1\\
m & -m'& m''
\end{array}
\right).
\label{ABC}
\end{align}

The recurrence relations for the Wigner 3-$j$ symbols are as follows: 
\begin{align}
\begin{array}{l}
-\sqrt{\left(l_{3} \mp s_{3}\right)\left(l_{3} \pm s_{3}+1\right)}\left(\begin{array}{ccc}
l_{1} & l_{2} & l_{3} \\
s_{1} & s_{2} & s_{3} \pm 1
\end{array}\right)\\
\quad=\sqrt{\left(l_{1} \mp s_{1}\right)\left(l_{1} \pm s_{1}+1\right)}\left(\begin{array}{ccc}
l_{1} & l_{2} & l_{3} \\
s_{1} \pm 1 & s_{2} & s_{3}
\end{array}\right)\\
\qquad+\sqrt{\left(l_{2} \mp s_{2}\right)\left(l_{2} \pm s_{2}+1\right)}\left(\begin{array}{ccc}
l_{1} & l_{2} & l_{3} \\
s_{1} & s_{2} \pm 1 & s_{3}
\end{array}\right)
\end{array}
\end{align}

For the standard rotational Killing vectors about the $x, y$ and $z$ axes given by
\begin{align}
L_x&=\sin \varphi \partial_{\theta}+\cot \theta \cos \varphi \partial_{\varphi}, \\
L_y&=\cos \varphi \partial_{\theta}-\cot \theta \sin \varphi \partial_{\varphi}, \\
L_z &=\partial_{\varphi}.
\end{align}
the computation yields
\begin{align}
\bar L_1 = &- L_x \cdot\bar m=\frac{1}{\sqrt{2}}( i \cos \theta \cos \varphi-\sin \varphi)
= i \sqrt{\frac{2\pi}{3}}\big (\,{}_{-1} Y_{1-1}- \,{}_{-1}Y_{11}\big), \\
\bar L_2=&  - L_y  \cdot\bar m =\frac{1}{\sqrt{2}}(i \cos \theta \sin \varphi+\cos \varphi )
= -\sqrt{\frac{2\pi}{3}}\big (\,{}_{-1} Y_{11}+ \,{}_{-1}Y_{1-1} ) ,\\
\bar L_3=& - L_z  \cdot \bar m =\frac{i}{\sqrt{2}} \sin \theta=-2 i \sqrt{\frac{ \pi}{3}}\,{}_{-1}Y_{10}.
\end{align}

Using these results in \eqref{ABC}, after~some algebra, we obtain
\begin{align}
c^x_{lml'm'} =~&i(-1)^{m' } \sqrt{\frac{(2 l+1)(2 l'+1)}{4}}
\bigg\{
3\sqrt{(l'+2)(l'-1)}
\left(\begin{array}{ccc}
l & l' & 1\\
-2 & 1 & 1
\end{array}\right) \nonumber \\
 & +\sqrt{(l+3)(l-2)}
\left(\begin{array}{ccc}
l & l' & 1\\
-3 & 2 & 1
\end{array}\right) \bigg\}
 \bigg\{
 \left(\begin{array}{ccc}
l & l' & 1\\
m & -m'& -1
\end{array}
\right)  -   \left(\begin{array}{ccc}
l & l' & 1\\
m & -m'& 1
\end{array}
\right) \bigg\},
\label{AB1}
\end{align}
\begin{align}
c^y_{lml'm'} =	&(-1)^{m' +1} \sqrt{\frac{(2 l+1)(2 l'+1)}{4}}\bigg\{
3\sqrt{(l'+2)(l'-1)}
\left(\begin{array}{ccc}
l & l' & 1\\
-2 & 1 & 1
\end{array}\right) \nonumber \\
 & +\sqrt{(l+3)(l-2)}
\left(\begin{array}{ccc}
l & l' & 1\\
-3 & 2 & 1
\end{array}\right) \bigg\}
 \bigg\{
 \left(\begin{array}{ccc}
l & l' & 1\\
m & -m'& 1
\end{array}
\right)  +   \left(\begin{array}{ccc}
l & l' & 1\\
m & -m'& -1
\end{array}
\right) \bigg\},
\label{AB2}
\end{align}
\begin{align}
c^z_{lml'm'} =	&i(-1)^{m'+1 }\sqrt{(2 l+1)(2 l'+1)}
\bigg\{
3\sqrt{(l'+2)(l'-1)}
\left(\begin{array}{ccc}
l & l' & 1\\
-2 & 1 & 1
\end{array}\right) \nonumber \\
 & +\sqrt{(l+3)(l-2)}
\left(\begin{array}{ccc}
l & l' & 1\\
-3 & 2 & 1
\end{array}\right) \bigg\}
 \left(\begin{array}{ccc}
l & l' & 1\\
m & -m'& 0
\end{array}
\right). 
\label{AB3}
\end{align}
\end{widetext}
In particular, for $l=2$, the~computation of the angular momentum $z$-component loss yields
\begin{align}
	\dot J_z =   &\frac{1}{4\pi}  \Im\sum^{m=2}_{m = -2}m h_{2m} \dot{\bar h}_{2m}.
\end{align}

In the final equation, we derive the momenta associated with the polarizations {$h$} of gravitational waves, based on gauge-invariant perturbations, as~outlined by Ruiz~\cite{52}, Thorne~\cite{25}, and Lousto~\cite{47}.



\bibliography{MDPI_Bibliography}

\end{document}